\documentclass[
    ,final            % use final for the camera ready runs
  ]
  {aipproc}

\layoutstyle{6x9}
\usepackage{amssymb,wrapfig}
\usepackage[labelsep=none]{caption}
\usepackage[absolute,overlay]{textpos}

\begin{document}

\title{CP Violation in SUSY Cascade Decay Chains\\ at the LHC}

\classification{12.60.Jv}
\keywords      {supersymmetric models, CP violation, LHC}

\author{G. Moortgat-Pick}{
  address={IPPP, University of Durham, Durham DH1 3LE, United Kingdom}
}

\author{K. Rolbiecki}{
  address={IPPP, University of Durham, Durham DH1 3LE, United Kingdom}
}

\author{J. Tattersall}{
  address={IPPP, University of Durham, Durham DH1 3LE, United Kingdom}
}

\author{P. Wienemann}{
  address={Department of Physics, University of Bonn, Nussallee 12, D-53115 Bonn, Germany}
}

\begin{abstract}
We discuss the potential to observe effects of CP violation in squark decay chains at the LHC. As the CP-violating observable we use the asymmetry composed by triple products of final state momenta. Extending methods for momentum reconstruction we show that there are good prospects for observation of these effects at the LHC. Finally, we include the main experimental factors and discuss the expected sensitivity.
\end{abstract}

\maketitle

%%%%%%%%%%%%%%%%%%%%%%%%%%%%%%%%%%%%%%%%%%%%
%% MAINMATTER
%%%%%%%%%%%%%%%%%%%%%%%%%%%%%%%%%%%%%%%%%%%%
\vspace*{-0.7cm}
\section{Introduction}

The search for supersymmetry (SUSY) is one of the main goals of present and future colliders since it is one of the best motivated extensions of the Standard Model (SM). An important feature of SUSY models is the possibility of introducing many new sources of CP violation. A careful analysis of new CP-violating effects will be required and in the following we discuss the example in the Minimal Supersymmetric Standard Model.

\begin{figure}
\resizebox{\textwidth}{!}
{\includegraphics[angle=270]{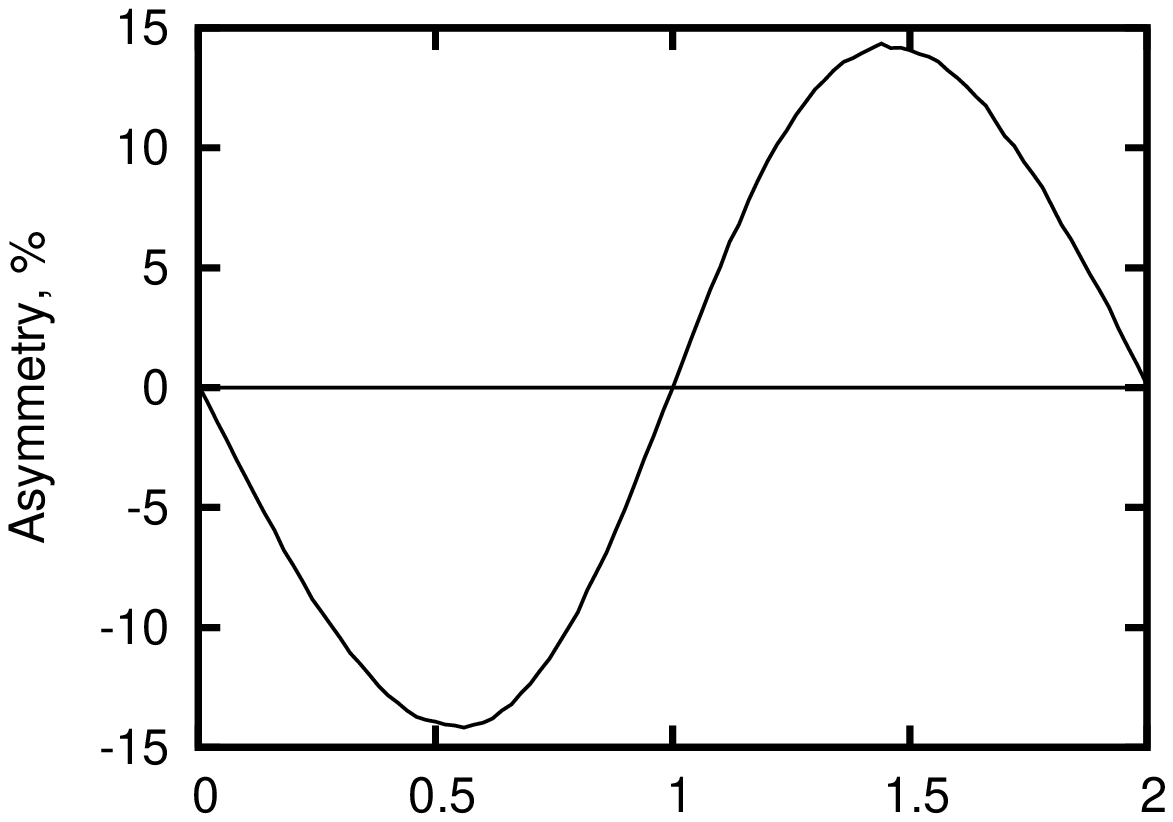}\includegraphics[angle=270]{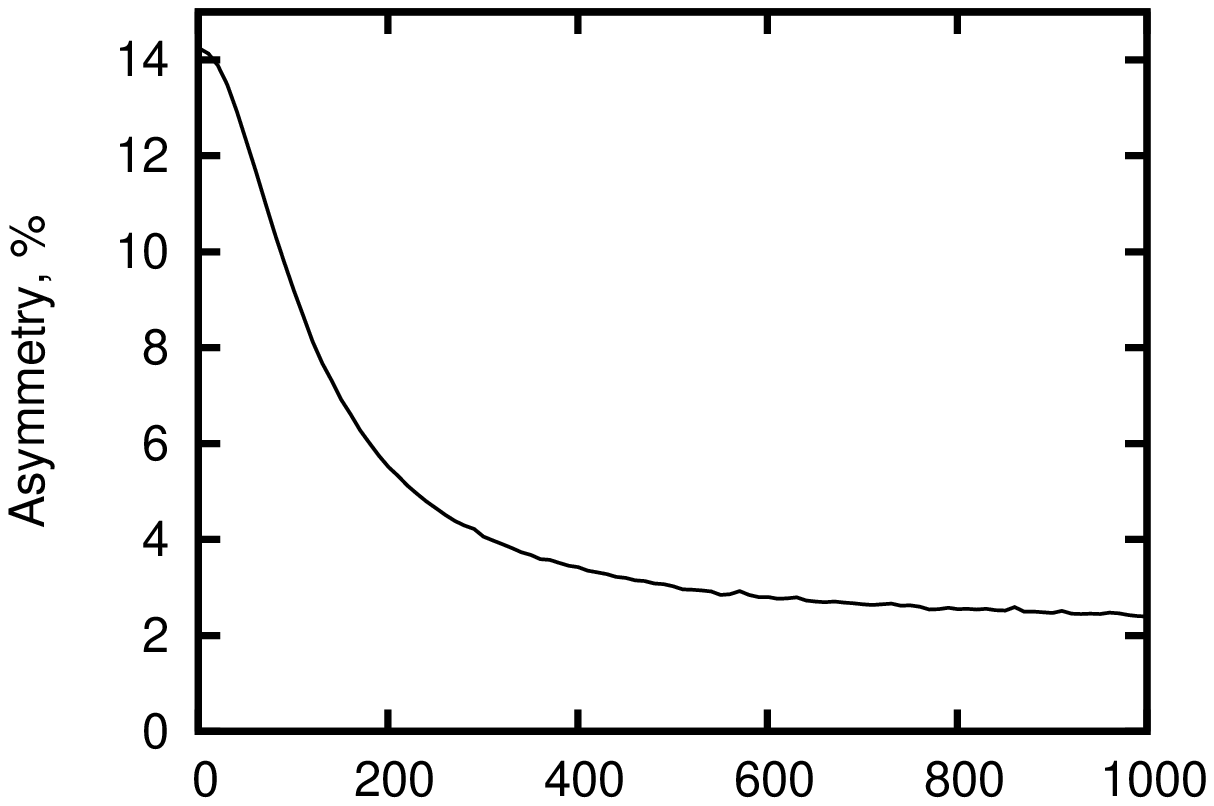}}
\caption{\textbf{(a):} The asymmetry $\mathcal{A}_T$, Eq.~(\ref{eq:asy}), in the rest frame of $\tilde{\chi}^0_2$ as a function of $\phi_1$. \textbf{(b):} The asymmetry $\mathcal{A}_T$, Eq.~(\ref{eq:asy}), in the laboratory frame as a function of the squark momentum, $|\vec{p}_{\tilde{q}}|$. \label{fig:asymmetry}}
\end{figure}

In Ref.~\cite{MoortgatPick:2009jy} we focus our attention on the possibility of observing CP-violating effects in squark decay chains at the Large Hadron Collider (LHC)
\begin{equation}\label{eq:chain}
\tilde{q} \to \tilde{\chi}_2^0 + q \to \tilde{\chi}_1^0 \ell^+ \ell^- + q \; , 
\end{equation}
where we have the three-body leptonic decay of the neutralino $\tilde{\chi}^0_2$ in the last step.
The main source of CP-violation is the phase of the bino mass parameter $M_1 = |M_1| \mathrm{e}^{\mathrm{i} \phi_1}$. As an observable~\cite{Choi:2005gt} we choose the T$_N$-odd triple product of momenta of the final state particles
\begin{equation}\label{eq:triple}
\mathcal{T}= \vec{p}_q \cdot (\vec{p}_{\ell^+} \times \vec{p}_{\ell^-})\; .
\end{equation}
Using this triple product one can construct a CP-odd asymmetry
\begin{equation}\label{eq:asy}
\mathcal{A}_T = \frac{N_{\mathcal{T}_+}-N_{\mathcal{T}_-}}{N_{\mathcal{T}_+}+N_{\mathcal{T}_-}}\; ,
\end{equation}
where $N_{\mathcal{T}_+}$ ($N_{\mathcal{T}_-}$) are the numbers of events for which
$\mathcal{T}$ is positive (negative), see also~\cite{Bartl:2004jj}.
At the parton level, in the neutralino $\tilde{\chi}_2^0$ rest frame, the asymmetry can be as large as $15\%$, cf.\ Fig.~\ref{fig:asymmetry}a. However, particles produced at the LHC get large, undetermined boosts that are a consequence of the internal proton structure. Due to these boosts the asymmetry is strongly diluted as can be seen in Fig.~\ref{fig:asymmetry}b. This makes the observation of CP-violating effects difficult~\cite{Ellis:2008hq}.

We show that a very useful tool in such an analysis is the reconstruction of momenta of all the particles involved in the process, including those escaping detection~\cite{MoortgatPick:2009jy}. Using this technique one can recover the large asymmetry present at parton level greatly increasing the discovery potential. 

\begin{textblock}{5}(2.3,2)
\small {(a)}
\end{textblock}

\begin{textblock}{5}(7.8,2)
\small {(b)}
\end{textblock}

\begin{textblock}{5}(2.4,10.6)
\small {(a)}
\end{textblock}

\begin{textblock}{5}(7.8,10.6)
\small {(b)}
\end{textblock}

\vspace*{-0.cm}
\section{CP asymmetry without momentum reconstruction}

\begin{figure}[!b]
%\resizebox{\textwidth}{!}
{\hskip -5pt \includegraphics[width=190pt,angle=0]{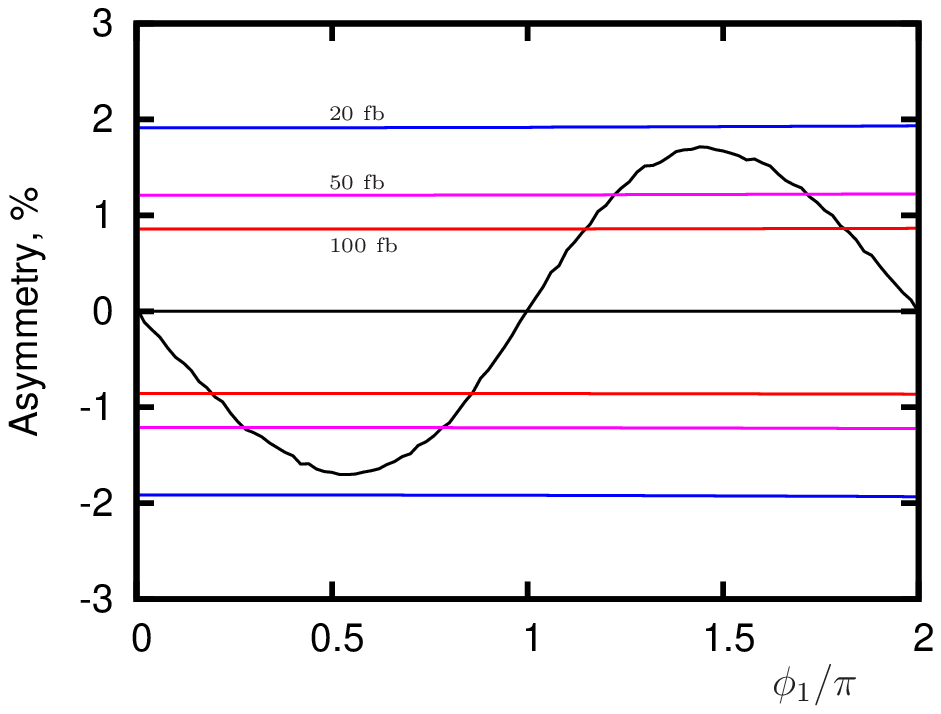}\hskip 0.5cm \includegraphics[width=190pt,angle=0]{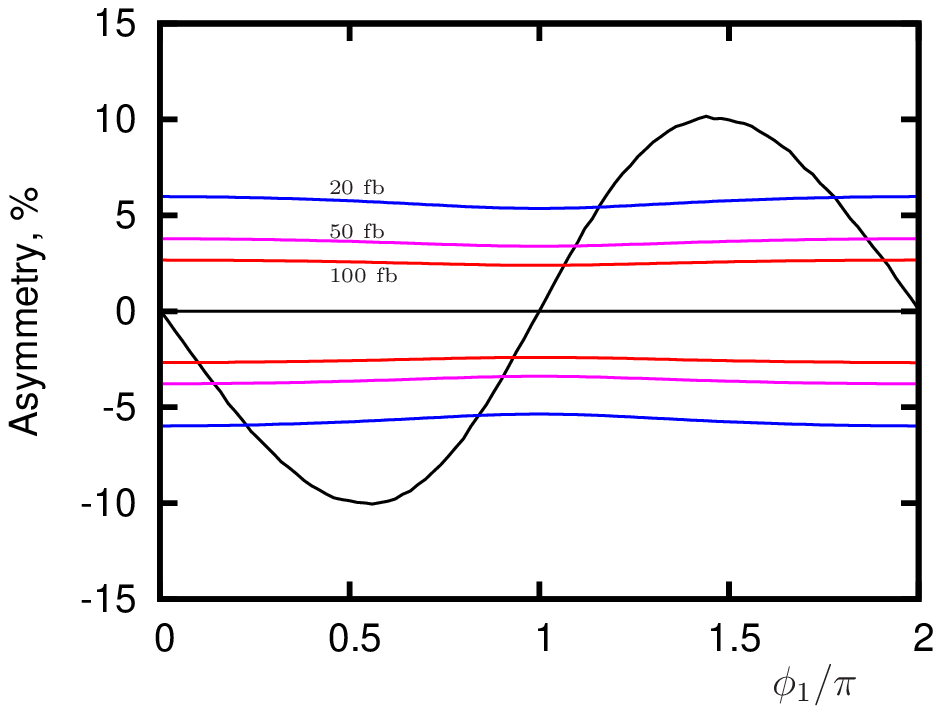}}
\caption{The asymmetry $\mathcal{A}_T$ at the LHC with PDFs included in the analysis: (a) before momentum reconstruction and (b) after momentum reconstruction. The coloured lines show the size of the asymmetry needed for a $3\sigma$ observation at the given luminosity and for $\sqrt{s} = 14$~TeV.\label{fig:results}}
\end{figure}

First we study the behaviour of the asymmetry after the inclusion of parton density functions (PDFs). Our observable, Eq.~\eqref{eq:asy}, is now significantly reduced due to boosts compared with the asymmetry in the neutralino rest frame, where it is maximal, see Fig.~\ref{fig:results}a. This is because a boosted frame can make the momentum vector of the quark appear to come from the opposite side of the plane formed by $\ell^+$ and $\ell^-$, hence changing the sign of the triple product, Eq.~(\ref{eq:triple}). The other dilution factor that have to be taken into account are anti-squarks $\tilde{q}^*_L$ that will be produced along with squarks. The asymmetry due to anti-squarks has the opposite sign compared to squarks, hence having them in the same number would wash out the asymmetry completely. However, due to the valence quarks present in the colliding protons, squark production will dominate over anti-squark production and the asymmetry can be still observed~\cite{MoortgatPick:2009jy}.

Inclusion of PDFs and anti-squarks reduce the asymmetry by about factor of 8 in our case. The maximum asymmetry is now $|\mathcal{A}_T| = 1.7\%$. The total production cross section and the respective branching ratios for the decay chain Eq.~(\ref{eq:chain}) give the expected number of events. With that we can estimate the integrated luminosity required to observe the asymmetry at $3\sigma$ level, as shown in Fig.~\ref{fig:results}a. We note that for $\mathcal{L} = 100\ \mathrm{fb}^{-1}$ a large range of values for $\phi_1$ is covered. 

\vspace*{-0.1cm}
\section{CP asymmetry with momentum reconstruction}

\renewcommand{\thefigure}{}
\setlength{\columnsep}{5mm}
\begin{wrapfigure}[21]{r}{0.5\textwidth}\nonumber
\centering \includegraphics[width=0.45\textwidth,angle=0]{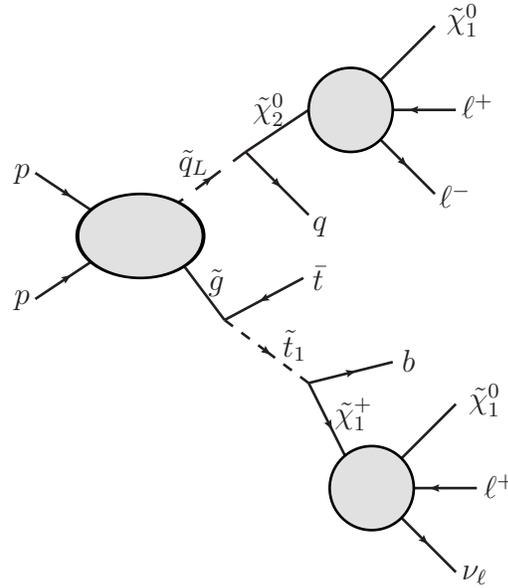}
\caption{\footnotesize \textbf{FIGURE 3.}\ \  The process studied for momentum reconstruction.}
\end{wrapfigure}
In order to overcome the dilution factor due to PDFs, we investigate the possibility of reconstructing the
momenta of the invisible particles in the process~\cite{MoortgatPick:2009jy,Kawagoe:2004rz}. The squark decay chain offers too little kinematical constraints to perform the reconstruction, hence we consider associated production of squark and gluino followed by the decay of the gluino  
\begin{equation}\label{eq:gluinodecay}
 \tilde{g} \to \tilde{t}_1\; \bar{t} \to \tilde{\chi}^+_1\;  b\;  \bar{t} \to \tilde{\chi}^0_1\; \ell^+\; \nu_{\ell}\;  b\;  \bar{t}\; .
\end{equation}
with the three-body decay of chargino in the last step, see Fig.~3.

For this process one can formulate 6 invariant-mass conditions for the intermediate particles and the final state LSP from the squark decay chain. Together with two equations involving missing transverse momentum this gives 8 equations (6 linear and 2 quadratic). We have 8 unknowns (the components of the four-momenta of the invisible final state particles) and can therefore solve the system~\cite{MoortgatPick:2009jy}. In principle such a system of equations can give up to four real solutions. Since we do not have any additional constraints to pick the correct solution, we include only those events that give the same sign for the triple product, Eq.~(\ref{eq:triple}), for all real solutions. This guarantees that we take the correct sign for the triple product for the calculation of the asymmetry.

Inclusion of all the branching ratios for the decay chains, Eq.~(\ref{eq:chain}) and (\ref{eq:gluinodecay}), and the requirement of having only the same sign triple products for all solutions significantly reduces the number of available events. However, it is now possible to reconstruct the momenta of the intermediate particles, in particular $\tilde{q}_L$ and $\tilde{\chi}^0_2$. This gives the possibility of calculating the triple product in the rest frame of the neutralino $\tilde{\chi}_2^0$ and restoring its maximal magnitude, as can be seen in Fig.~\ref{fig:results}b. The dilution, cf.\ Fig.~\ref{fig:asymmetry}a, is due to the anti-squark admixture but there is a significant improvement compared to the situation before the reconstruction, cf.\ Fig.~\ref{fig:results}a.

\begin{wrapfigure}{r}{0.5\textwidth}
\vspace*{-0.cm}\centering \includegraphics[width=190pt,angle=0]{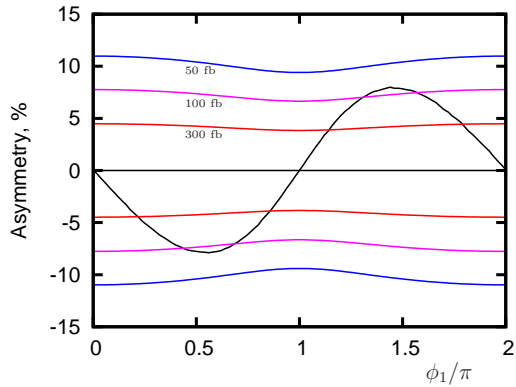}\vspace*{-0.1cm}
\caption{\footnotesize \textbf{FIGURE 4.}\ \  The asymmetry $\mathcal{A}_T$ at the LHC after momentum reconstruction with cuts and momentum smearing. The coloured lines show the size of the asymmetry needed for a $3\sigma$ observation at the given luminosity  and for $\sqrt{s} = 14$~TeV.}
\end{wrapfigure}
Finally, we take into account some of the experimental factors in our analysis. These include basic selection cuts and momentum smearing due to the finite detector resolution, see Ref.~\cite{MoortgatPick:2009jy} for details. Momentum smearing increases the dilution as we no longer correctly reproduce the rest frame of the neutralino $\tilde{\chi}^0_2$. Another consequence is a further reduction in the number of usable events. Nevertheless, after inclusion of these effects we are still able to observe the asymmetry, as can be seen in Fig.~4. With an integrated luminosity of $\mathcal{L} = 300\ \mathrm{fb}^{-1}$ we expect to have sensitivity to phases in the ranges $0.2\,\pi \lesssim \phi_{1} \lesssim 0.85\,\pi$ and $1.15\,\pi \lesssim \phi_{1} \lesssim 1.8\,\pi$ at the $3\sigma$ level.

\vspace*{-0.5cm}
\section{Conclusions}

In \cite{MoortgatPick:2009jy} we have studied how to observe CP-violating effects in SUSY cascade decay chains at the LHC. We have shown that by reconstructing the momenta of invisible particles one can get access to precision observables that are sensitive to CP-violating phases, like triple products of momenta. Measurement of CP phases is important not only to shed a light on the origin of the matter-antimatter asymmetry but also to direct future searches at a linear collider.

%%%%%%%%%%%%%%%%%%%%%%%%%%%%%%%%%%%%%%%%%%%%%%%%
%% BACKMATTER
%%%%%%%%%%%%%%%%%%%%%%%%%%%%%%%%%%%%%%%%%%%%%%%%

\begin{theacknowledgments}

KR is supported by the EU Network MRTN-CT-2006-035505 (HEPTools). JT is supported by the UK Science and Technology Facilities Council (STFC).
\end{theacknowledgments}

\bibliographystyle{aipproc}

\end{document}